\pdfoutput=1
\documentclass[letter]{aa} 

\usepackage{graphicx}
\usepackage{txfonts}
\usepackage[allcolors=blue]{hyperref}
\newcommand{\xv}{\mathbf{x}}

\newcommand{\kv}{\mathbf{k}}

\newcommand{\pv}{\mathbf{p}}
\newcommand{\qv}{\mathbf{q}}

\newcommand{\delobs}{\delta_\mathrm{obs}}

\newcommand{\pobs}{P_\mathrm{obs}}
\newcommand{\Bobs}{B_\mathrm{obs}}
\newcommand{\dq}{\frac{\mathrm{d}^3 q}{(2\pi)^3}}
\newcommand{\Dp}{\frac{\mathrm{d}^3 p}{(2\pi)^3}}

\newcommand{\be}{\begin{equation}}
\newcommand{\ee}{\end{equation}}
\newcommand{\bea}{\begin{eqnarray}}
\newcommand{\eea}{\end{eqnarray}}
\newcommand{\bdm}{\begin{displaymath}}
\newcommand{\edm}{\end{displaymath}}

\newcommand{\de}{\mathrm{d}}

\newcommand{\Mpc}{\, h^{-1} \, {\rm Mpc}}
\newcommand{\cMpc}{\, h^{-3} \, {\rm Mpc}^3}

\newcommand{\kMpc}{\, h \, {\rm Mpc}^{-1}}

\newcommand{\omegam}{\Omega_\text{m}}

\newcommand{\omegab}{\Omega_\text{b}}

\usepackage{todonotes}

\defcitealias{FKP1994}{FKP}
\usepackage{ulem}

\begin{document}

   \title{Window function convolution with deep neural network models}

   \subtitle{}

   \author{D. Alkhanishvili
          \inst{1}\fnmsep\thanks{Member of the International Max Planck Research School (IMPRS) for Astronomy and Astrophysics at the Universities of Bonn
and Cologne}\fnmsep\thanks{\email{daalkh@astro.uni-bonn.de}}
          \and
          C. Porciani\inst{1}
          \and
          E. Sefusatti\inst{2,3,4}
          }

   \institute{Argelander Institut für Astronomie der Universität Bonn, Auf dem Hügel 71, 53121 Bonn, Germany
         \and
             Istituto Nazionale di Astrofisica, Osservatorio Astronomico di Trieste, via Tiepolo 11, 34143 Trieste,
Italy
        \and
        Istituto Nazionale di Fisica Nucleare, Sezione di Trieste, via Valerio 2, 34127 Trieste, Italy
        \and
        Institute for Fundamental Physics of the Universe, Via Beirut 2, 34151 Trieste, Italy
             }

   \date{Received October 7, 2022; accepted December 6, 2022}

 
  \abstract
   {Traditional estimators of the galaxy power spectrum and bispectrum are sensitive to the
survey geometry. They yield spectra that differ from the true
underlying signal since they are convolved with the window function of the survey.
For the current and future generations of experiments, 
this bias is statistically significant on large scales.
It is thus imperative that the effect of the window function on the summary statistics
of the galaxy distribution is accurately modelled. Moreover, this operation must be
computationally efficient in order to allow sampling posterior probabilities 
while performing Bayesian estimation of the cosmological parameters.
In order to satisfy these requirements, 
we built a deep neural network model that emulates the convolution with the window
function, and we  show that it provides fast and accurate predictions.
We trained (tested) the network using a suite of 2000 (200) cosmological models
within the cold dark matter scenario, and demonstrate that
its performance is agnostic to the precise values of the cosmological
parameters. In all cases, the deep neural network provides models for the power spectra and the bispectrum that are accurate to better than 0.1 per cent
on a timescale of 10 $\mu$s.}


   \maketitle

%

\section{Introduction}

The most common approach to extract cosmological information from a galaxy redshift survey involves measuring the power spectrum and/or the bispectrum of the galaxy distribution.
In the majority of cases, the spectra are derived using traditional estimators
\citep{Yamamoto+2006,Bianchi+2015,Scoccimarro+2015}
based on the ideas originally introduced by~\citet[][hereafter \citetalias{FKP1994}]{FKP1994}. One drawback of this method is that
the survey geometry leaves an imprint on the measured spectra,
which is difficult to model.
The observed galaxy overdensity field $\delobs(\xv)$ does not coincide
with the actual fluctuations $\delta(\xv)$. The reasons are twofold. First, galaxy surveys cover only finite sections of our past light cone. Second, tracers of the large-scale structure need to be weighted based on the selection criteria of the survey. 
In compact form we write $\delobs(\xv)=W(\xv)\,\delta(\xv)$, where
$W(\xv)$ denotes the window function of the survey.
It follows that 
the observed power spectrum and the underlying true power $P(k)$
satisfy the relation~\citep[][\citetalias{FKP1994}]{PeacockNicholson1991}

\begin{equation}
\label{eq:power_conv}
    \pobs (\kv) = \int |\Tilde{W}_2(\kv-\qv)|^2 \,P(\qv)\,\dq\,,
\end{equation}
where $\Tilde{W}_n(\kv)$ denotes the Fourier transform of the function $W(\xv)$ normalised
such that 
\begin{align}
\label{eq:w(k)}
    \Tilde{W}_n(\kv) = \frac{\int W(\xv)\, \mathrm{e}^{i\,\kv\cdot\xv}\, \de^3 x}{ \left\{\int [W(\xv)]^n\,\de^3 x \right\}^{1/n} }\, .
\end{align}
Similarly, for the bispectrum we obtain
\begin{align}
\label{eq:bisp_conv}
\nonumber
    \Bobs (\kv_1,\kv_2,\kv_3)=\int &\tilde{W}_3(\kv_1-\qv)\, \tilde{W}_3(\kv_2-\pv)\,\tilde{W}_3(\kv_3+\qv+\pv) \\ & B(\qv,\pv,-\qv-\pv) \,\dq\, \Dp \, .
\end{align}

The convolution in Eq.~(\ref{eq:power_conv}) mixes Fourier modes with different wavevectors and modifies the power\footnote{
A different class of estimators based on pixelised maps of the galaxy density
is immune to this problem and directly provides noisy measurements of the unwindowed spectra (see e.g. the so-called quadratic estimators for the power spectrum in \citealt{Tegmark_1998} and \citealt{Philcox+2021} and the cubic estimator for the bispectrum in \citealt{PhilcoxWindow2021}). However, these estimators are sub-optimal on small scales. Moreover, at the moment, there is no such estimator for the anisotropic bispectrum in redshift space.} significantly on large scales.
Since the survey window is generally not spherically symmetric, it also creates an anisotropic signal in addition to redshift-space distortions and the Alcock-Paczynski effect.
These consequences need to be accounted for in order to fit theoretical models to the observational data, in particular when trying to constrain the level of  
primordial non-Gaussianity \citep[e.g.][]{Castorina+20}
or general relativistic effects \citep[e.g.][]{Elkhashab+22}.
Two approaches are possible: by  trying to remove the effect from the data \citep[e.g.][]{Sato+2011} or accounting for the window in the models \citep[e.g.][]{deLaixStarkman+98,Percival+2001,Ross+2013}.
This second line of attack is much more popular:
starting from an estimate for the window function, a model for $\pobs (\kv)$ is obtained by solving the convolution integral numerically.
Developing numerical procedures for 
fast likelihood evaluation is pivotal in multivariate Bayesian inference.
For this reason, 
\cite{Blake+13} reformulated the convolution integrals as matrix multiplications
and made use of pre-computed `mixing matrices' to evaluate the impact of the survey window
on the power spectra averaged within wavector bins.
A computationally efficient way to evaluate the effect of the window function on the multipole moments of the power spectrum in the distant-observer approximation is presented by \cite{Wilson2017} and generalised by \cite{Beutler+17} to the local plane-parallel case (in which the line of sight varies with the galaxy pair). In this approach the convolution is cast in terms of a sequence of one-dimensional Hankel transforms that are performed using the FFTlog algorithm~\citep{Hamilton2000}. 
The key idea is to
compress the information about the window function into a finite number of multipole moments of its autocorrelation function \citep[see also][]{Beutler+14}.
Further extensions account for wide-angle effects \citep[e.g.][]{Castorina-White+2018, Beutler+19}.

Evaluating the impact of the survey window on the bispectrum has only recently received attention in the literature.  From a computational perspective,
performing the six-dimensional convolution integral in Eq.~(\ref{eq:bisp_conv}) is a  challenging task that cannot form the basis of a toolbox for Bayesian inference.
It is thus necessary to develop faster techniques.
Inspired by perturbation theory at leading order, \cite{GilMarin2015} proposed an approximation where the monopole moment of the convolved bispectrum is given by the linear superposition of products of two convolved power spectra given by Eq.~\eqref{eq:power_conv}. Although this approximation is accurate enough for the BOSS survey (barring squeezed
triangular configurations, 
which are   excluded from the analysis by \citealt{GilMarin2015}),
it would likely introduce severe biases 
in the analysis of
the next generation of wide and deep surveys such as Euclid~\citep{Euclid2011} or DESI~\citep{DESI2016}, which will provide measurements with much smaller statistical uncertainties~\citep[see e.g.][]{Yankelevich2019}. 
\cite{Sugiyama2019} introduced a new bispectrum estimator based on the 
tri-polar spherical harmonic decomposition with zero total angular momentum
and showed, in this case, that it is possible to compute the models 
for the convolved bispectrum following a FFT-based approach.
The issue of developing a similar method for more traditional estimators
of the bispectrum multipoles \citep{Scoccimarro+2015}
has been recently addressed by
\cite{Pardede2022}, who derived an expression
based on two-dimensional
Hankel transforms that can be computed using the 2D-FFTlog method \citep{Fang+2020}.
In this case the survey window is described in terms of 
the multipoles of its three-point correlation function.
Developing optimal estimators for these quantities is still an open problem.

 In this letter we propose   employing deep learning as a method to
 compute the impact of the survey window function on theoretical models for the power spectrum and bispectrum.
 Specifically, we use a deep neural network (DNN) to approximate the mapping
 from the unconvolved to the convolved spectra. 
 This technique allows us to consider multiple cosmological models, while drastically reducing computer-memory demands and the wall-clock time of computation with respect to performing the convolution integrals numerically. All these features are key for building
 efficient Bayesian inference samplers and determining the posterior distribution of cosmological parameters. 
The structure of the letter is as follows. In Sect. \ref{sec:methods} 
we briefly describe the architecture of our DNN models and introduce 
the data sets we employ for training and testing them. Our results are presented in Sect. \ref{sec:results}. We draw our conclusions in Sect. \ref{sec:conclusions}.

\section{Methods}
\label{sec:methods}

\subsection{Philosophy and goals}
It is well known that artificial neural networks are able to approximate any arbitrary continuous function of real variables~\citep{Cybenko1989,WHITE1990,HORNIK1991}. 
They learn how to map some inputs (features, in machine learning jargon) to outputs (labels) from examples in a training data set. 
The training process consists of fitting the parameters of the machine (weights and biases of the neurons) by minimising
a loss function that quantifies how good  the prediction is with respect to the correct result. 
After the training the accuracy of the model is determined using the testing data. 

In our applications the features that form the input of the DNN are the spectra
$P(\kv)$ and $B(\kv_1,\kv_2,\kv_3)$ evaluated at specific sets of wavevectors.
Different options are available when choosing these sets. For instance, we could use many closely separated wavevectors around the output configurations. In this case the DNN would learn how the convolution integrals mix the contributions coming from different configurations. At the opposite extreme, we could consider inputs and outputs evaluated for the very same set of configurations, so that, in some sense, the DNN model also interpolates among the sparser inputs. 
We opted for this second approach, which is more conducive to a simpler machine learning set-up: choosing a smaller size of features
requires fewer model parameters to be tuned, and the trained model is evaluated more quickly. The only implicit assumption here is that
the input power spectrum is smooth  between the sampled configurations.
In our implementation
the machine learns to predict the functions
\begin{equation}
\label{eq:ratio_P}
    R_P(\kv) = \frac{\pobs(\kv)}{P(\kv)}
    \ \ \ \mathrm{and} \ \ \ 
    R_B(\kv_1,\kv_2,\kv_3) = \frac{\Bobs(\kv_1,\kv_2,\kv_3)}{B(\kv_1,\kv_2,\kv_3)}
\end{equation}
evaluated at the same arguments of the input.

\begin{figure}

        \includegraphics[width=\columnwidth]{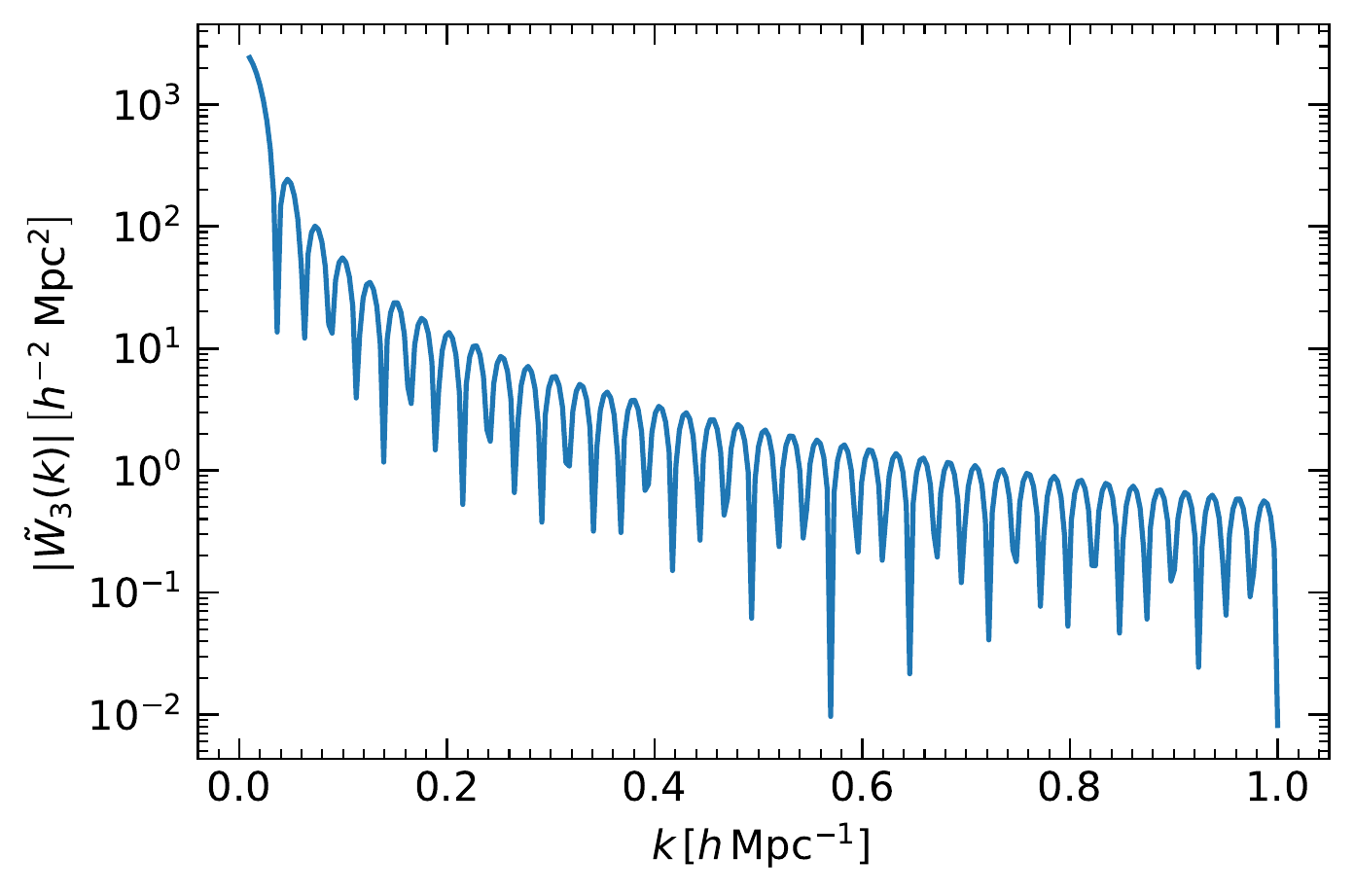}
    \caption{Top-hat window function with $V=200^3\,\cMpc$.}
    \label{fig:top_hat}
\end{figure}

Since this letter is about giving a proof of concept,
for simplicity we only predict the effect of window function on the linear matter power spectrum and the so-called tree-level bispectrum, which can be trivially computed using the linear power spectrum~\citep[e.g.][]{Fry+84}, neglecting redshift-space distortions in both cases.
Moreover, as an example, we consider a spherically symmetric top-hat window function which
assumes the value of one for distances smaller than the radius $R$ and zero otherwise.
In Fourier space this corresponds to
\begin{equation}
    \label{eq:tophat_fourier}
    \tilde{W}_n(k) = \frac{4\pi\,(k R)^2\,j_1(k R)}{k^3\,V^{1/n}} \, ,
\end{equation}
where the symbol $j_\nu(x)$ denotes the spherical Bessel functions of the first kind and $V=4\pi\,R^3/3$
is the comoving volume enclosed by the window function. 
Basically, $\tilde{W}_n(k)$ rapidly oscillates 
(see Fig.~\ref{fig:top_hat}) which makes it challenging to numerically compute the integrals in Eqs. \eqref{eq:power_conv} and \eqref{eq:bisp_conv}. 
The oscillations are damped, and the main contribution to the convolution comes 
from the first peak at $k=0$, which mixes Fourier modes within a shell of width $\Delta k\simeq R^{-1}$.
Given our assumptions, $R_P(k)$ and $R_B(k_1,k_2,k_3)$ only
depend on the modulus of the wavevectors.

\subsection{Deep learning models}

Since the power spectrum is a smooth function of $k$, we adopt
the convolutional neural network (CNN) architecture to model $R_P(k)$. The first layer of the network applies a convolution to the input with 16 trainable filters (kernel size 3) and a rectified linear unit (ReLU) activation function, defined as $f(x)=\max (0,x)$. This is followed by a dropout layer with a rate of 0.5, which acts as a regulariser and prevents overfitting~\citep{Goodfellow-et-al-2016}. The final layer is a dense one in which the number of neurons matches the length of the output data vector. Since the convolved power spectrum must be positive, the last layer is processed through a softplus activation function of the form $f(x)=\ln(1+\mathrm{e}^x)$. 

For the bispectrum, we opt for a different CNN architecture.
We chose this  because we organise the data in a one-dimensional array 
where each entry corresponds to a different triangular configuration. 
As a consequence, the sequence of data is not necessarily smooth and
we prefer to use a model that  can detect features on multiple scales.
The network we chose is based on the U-Net architecture~\citep{Ronneberger2015}, 
which consists of a contracting path (encoder) followed by an expansive path (decoder).
The former combines convolutional and pooling layers to down-sample
the original data, and thus builds a compressed representation of them.
The latter decompresses the compact representation to construct
an output of the desired size.
We include two  down-sampling and two up-sampling steps,
followed by a dropout layer with a rate of 0.5 
and a dense layer 
in which
the number of neurons matches the length of the output data vector.

For building and training the neural networks, we use the Keras library~\citep{chollet2015keras} under the TensorFlow framework~\citep{tensorflow2015-whitepaper}. During the training phase, the parameters of the machine are adjusted to minimise a loss function $L$, which we identify with the
mean absolute error (MAE)
    $L = N^{-1}\,\sum_{i=1}^N |y^\mathrm{pred}_i/y^\mathrm{truth}_i-1|$,
where $y^\mathrm{pred}_i$ denotes the DNN prediction, $y^\mathrm{truth}_i$ is the corresponding item in the training data set, and $N$ is the number of entries in these data vectors. 
The minimisation of this loss function 
(one of the most popular choices for regression problems)
is controlled by the Adam Optimizer~\citep[based on the stochastic gradient descent method,][]{AdamPaper2014} with an initial learning rate of $0.001$, which is reduced as the training progresses using the inverse time decay schedule of TensorFlow. 
In order to prevent overfitting,
five per cent of the training data are set aside as a validation set (i.e. these data are not used to fit the network parameters), and training is automatically stopped early if the validation loss starts to increase. This step makes sure that the network predicts previously unseen data more accurately.

\subsection{Training and testing data sets}

\begin{table*}
        \centering
        \caption{Parameter spaces spanned by the training and testing data sets.}
        \label{tab:cosmo_params}
        \begin{tabular}{lccccc} 
                \hline 
                & \\[-9pt]
                 Data set & $\omegam$ & $\omegab$ & $h$ & $n_s$ & $\sigma_8$ \\ [3pt]
                \hline 
                & \\[-9pt]
                Training & [0.1,0.5] & [0.03,0.07] & [0.5,0.9] & [0.8,1.2] & [0.6,1.0] \\ [3pt]
                \hline
                & \\[-9pt]
                 Testing & [0.2,0.4] & [0.03,0.06] & [0.6,0.8] & [0.9,1.1] & [0.7,1.0]  \\ [3pt]
                \hline
        \end{tabular}
\end{table*}

We used the suite of 2000 linear power spectra in the Quijote database~\citep{QuijoteSims2019} to build the training set for our DNNs.
These spectra are obtained by sampling
five cosmological parameters on a Latin hypercube over the ranges defined in Table \ref{tab:cosmo_params}. 

For the power-spectrum analysis, we use 47 linearly spaced wavenumbers in the range
$[0.004,0.2]\,\kMpc$. The corresponding values for $\pobs(k)$
are computed using the 3D FFT method to evaluate Eq.~\eqref{eq:power_conv}
employing a top-hat window, which covers a comoving volume of $V=700^3\,\cMpc$ (i.e. $R=434.25\,\Mpc$). 

For the bispectrum we compute the tree-level expression from the linear power spectra
in the Quijote suite. We consider 564 triangular configurations
in which the three sides span the range $[0.01,0.2]\,\kMpc$.
In this case
the integration in Eq. \eqref{eq:bisp_conv} is carried out in six dimensions using the \textsc{Vegas} routine of the Cuba library~\citep{ Hahn2005}.
To facilitate the convergence of the numerical integrals, we consider a top-hat window with $V=200^3\,\cMpc$ ($R=124.07\,\Mpc$).

We generate the test data set by randomly generating 200 sets of cosmological parameters with Latin hypercube sampling. Since the Quijote database spans a broader region of parameter space compared with that allowed by current observational constraints, our test data are sampled within a narrower region mimicking the actual constraints from~\cite{Planck2018}; see the
bottom row of Table~\ref{tab:cosmo_params}.
For each set we compute the linear matter power spectrum using \textsc{camb}~\citep{Lewis:1999bs, Howlett:2002} and employ the same procedure used for the training data set to obtain the convolved power spectra and bispectra.

\section{Results}
\label{sec:results}
\begin{figure*}

        \includegraphics[width=\textwidth]{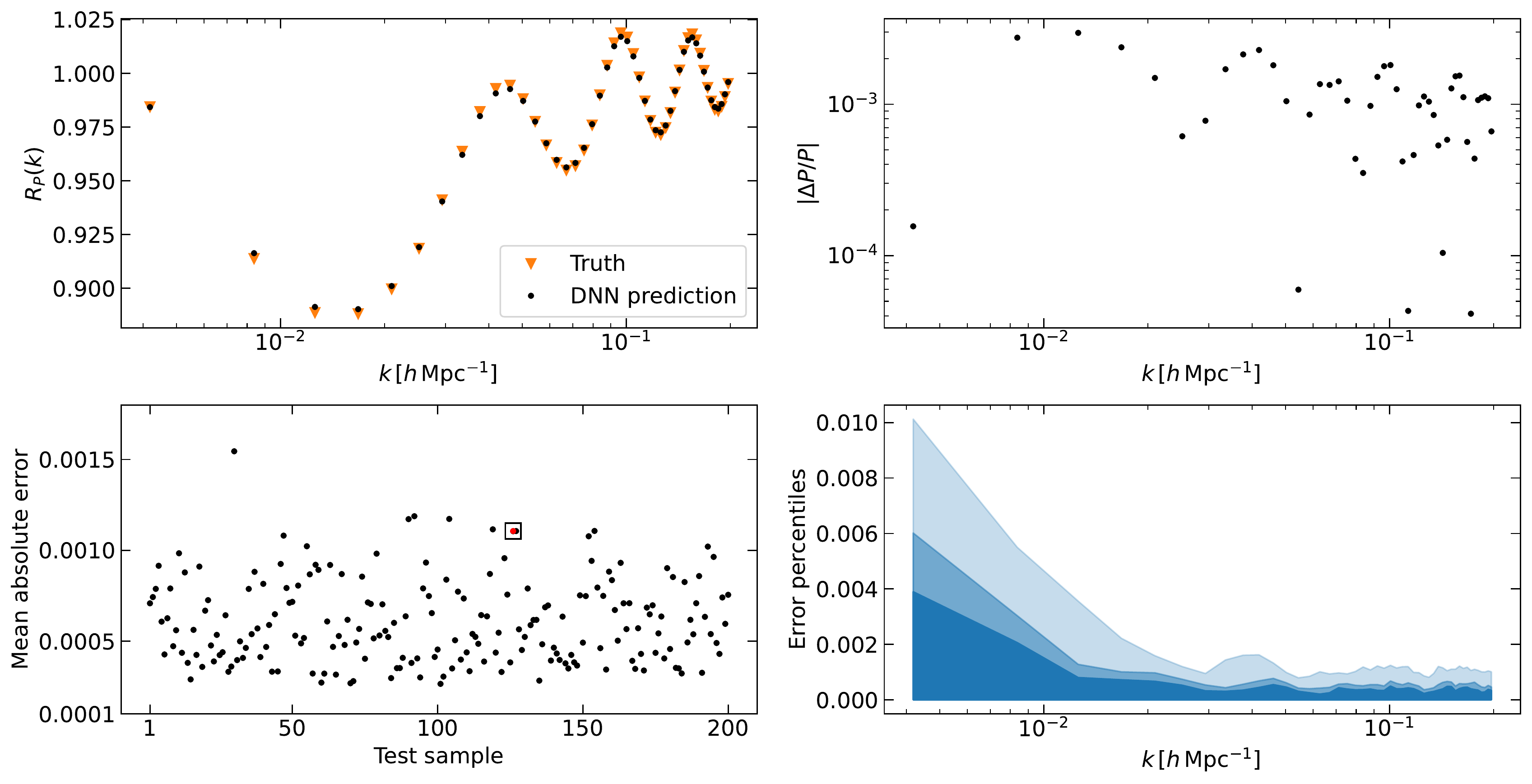}
    \caption{Accuracy of the trained DNN for the power spectrum convolution. Top left: Function $R_P(k)$ obtained with the convolution integral in
    Eq. \eqref{eq:power_conv} (orange triangles)  compared with the DNN model (black dots) for one test sample. Top right: Relative error of the DNN model in the same test sample used in the left panel. Bottom left: MAEs for all the test samples (the one used in the top panels is highlighted in red and surrounded by a square).  Bottom right: 50$\mathrm{th}$,
    68$\mathrm{th}$, and 95$\mathrm{th}$ error percentiles of the DNN model as a function of $k$. Generally,  the DNN model yields sub-per cent accuracy.}
    \label{fig:ps_result}
\end{figure*}

\begin{figure*}

        \includegraphics[width=\textwidth]{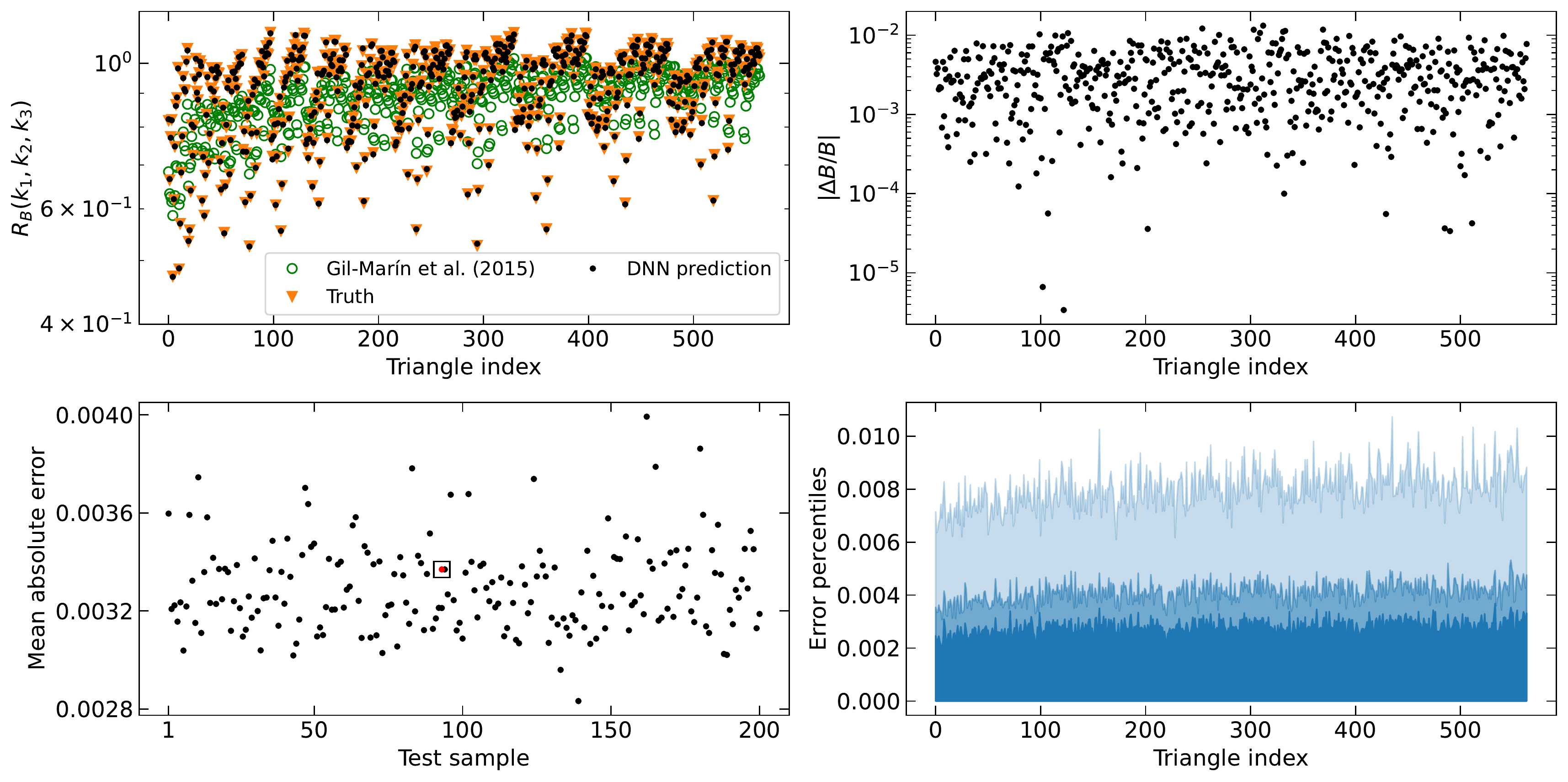}
    \caption{As in Fig.~\ref{fig:bs_result}, but for the bispectrum.  The triangular configurations in the top and bottom right panels satisfy the constraint $k_1 \geq k_2 \geq k_3$ and are ordered so that first $k_3$ increases (at fixed $k_1$ and $k_2$), then $k_2$ (at fixed $k_1$), and finally $k_1$. For reference, in the top left panel,  the ratio $R_B(k_1, k_2, k_3)$  is also plotted, computed according to the approximated
    method introduced by \cite{GilMarin2015} (green circles).}
    \label{fig:bs_result}
\end{figure*}

In the top panels of Fig.~\ref{fig:ps_result} we consider one of the test samples for
the power spectrum.
The orange triangles in the left panel show the function $R_P(k)$ computed using equation~\eqref{eq:power_conv}:
the convolution with the window function
flattens out the power spectrum on large scales and changes the amplitude of the baryonic acoustic oscillations by a few per cent. Although the window function considered here is arbitrary, 
similarly sized (and measurable) corrections are expected for
the next generation of galaxy redshift surveys 
\citep[see e.g. Fig. 6 in][]{Elkhashab+22},
which should
deliver per cent accuracy for the power spectrum.
The black dots indicate the output of the trained DNN model. The right panel shows the
relative error between the DNN prediction and the true signal, which is always smaller than 0.1 per cent.
To assess the overall performance of the DNN model, 
in the bottom panels of Fig.~\ref{fig:ps_result} we plot the 
MAEs for each test sample (left) and error percentiles over the test samples as a function of the wavenumber. The residual mean inaccuracy of the model is well below the  per cent level.

 The effect of the window function on the bispectrum is much more pronounced than for the power spectrum and the ratio $R_B(k_1,k_2,k_3)$ assumes values
 below 0.5 for some triangle configurations (top left panel in Fig.~\ref{fig:bs_result}). The DNN model predicts 
the corrections
accurately in all cases (top right and bottom panels)
and vastly outperforms 
the approximated method introduced by \cite{GilMarin2015},
which, for the compact survey volume considered here,
does not accurately reproduce the amplitude of the
convolved bispectrum
(green circles in the top left panel).

\section{Conclusions}
\label{sec:conclusions}

In this letter we employed a DNN model to predict the impact of the window function on the power spectrum and the bispectrum measured in a galaxy redshift survey. 
Overall, the trained DNN models show very promising results with sub-per cent MAEs
for all test samples (well below the statistical uncertainty expected from the next generation of surveys). These errors can be further reduced by increasing the size of the training data set. 

Our DNN model is meant as a proof of concept and, for this reason,
we made some simplifications in our study. First, we used the linear power spectrum and the tree-level bispectrum for matter fluctuations.
Second, we considered a top-hat window function with a fixed volume in which the
number density of tracers does not vary with the radial distance from the observer.
Although this is an ideal case, 
we do not see a reason why
 a DNN model should not be able to accurately predict the effect of more realistic survey masks, given an appropriately sized training sample.

It takes less than 10 microseconds to generate a complete sample
for either $R_P(k)$ or $R_B(k_1,k_2,k_3)$ with the trained DNN.
This is ideal for sampling posterior probabilities in Bayesian parameter estimation.
Our method can be straightforwardly generalised to the multipoles of the spectra, and
could also be combined with emulators
that make predictions for the true clustering signal (including galaxy biasing) based on perturbation theory \citep[e.g.][]{McCann+22,DeRose+22,Eggemeier+22}.
 Additional corrections due to binning the theory predictions in exactly the same way as done for the measurements (see e.g. Sect. 3.2 in \citealt{Oddo+20} and Sect. 4.1 in \citealt{Alkhanishvili+22}) can  
   be computed by suitably averaging the output of the DNN model
 or, more efficiently, can be accounted for in the model. Since in this letter we do not perform a Bayesian inference for cosmological parameters, we skipped this step when we   generated the training sample.

The bottleneck operation in the DNN approach is the creation of the training data set, which requires a significant time investment in the case of the bispectrum (in our case,
the calculation of the 2000 convolved bispectra with 64 processor cores took  approximately   one month of wall-clock time). This step can be sped up using massive parallelisation and, possibly, by relying on more computationally friendly formulations
of the convolution integral \citep[e.g.][]{Pardede2022}.
It is also conceivable
that using larger input vectors that densely sample the wavenumbers within
shells of size $\Delta k$ around the output configurations
might facilitate the task of the machine, and would thus help to reduce
the training data.
The time required to build the training set is not
a good reason to dismiss the DNN approach.
Even for the simple case of the isotropic bispectrum of matter-density fluctuations in real space,
sampling the posterior distribution of the five cosmological
parameters we considered would require many more than 2000 evaluations
of the window-convolved signal. Thus, using the DNN model would lead in any case to 
a notable speed up.
In any practical application, accounting for redshift-space distortions,
shot noise, and perturbative counterterms would substantially increase
the number of adjustable coefficients in the perturbative model for the bispectrum (and, correspondingly, the number of likelihood evaluations needed to constrain them from experimental data).
We thus conclude that using the DNN model would be advantageous
as long as the size of the necessary training set is substantially smaller 
than the number of the required likelihood evaluations in the Bayesian estimation
of the model parameters.

\begin{acknowledgements}
We thank Alexander Eggemeier and H\'ector Gil-Marin for useful discussions. D.A. acknowledges partial financial support by the Shota Rustaveli National Science Foundation of Georgia (GNSF) under the grant FR-19-498.
\end{acknowledgements}

%
%


\bibliographystyle{bibtex/aa}
\bibliography{manuscript} 

\end{document}